\documentclass[a4paper,11pt]{article}
\usepackage{pos}
\usepackage{physics}
\usepackage{graphicx}
\usepackage{float}

\newcommand{\Kahler}{K\"ahler }
\renewcommand{\O}{{\cal O}}
\newcommand{\cloop}{c_{\text{loop}}}
\newcommand{\V}{\mathcal{V}}
\newcommand{\N}{\mathcal{N}}

\title{Loop Blow-Up Inflation: a novel way to inflate with the \Kahler moduli}

\author*[a, b]{Luca Brunelli}

\affiliation[a]{Department of Physics and Astronomy, Bologna University \\
Via Irnerio 46, 40126, Bologna, Italy}

\affiliation[b]{INFN, Sezione di Bologna, viale Berti Pichat 6/2, 40127 Bologna, Italy}

\emailAdd{l.brunelli@unibo.it}

\abstract{The topic of this talk is a new inflationary model in the context of Type IIB string compactifications called \textit{Loop Blow-Up Inflation}, presented in \cite{Bansal:2024uzr}. The original \textit{Blow-Up Inflation} model \cite{Conlon:2005jm}, whose potential was purely non-perturbative, suffers from the $\eta$-problem, being sensitive to string-loop corrections. If these corrections are introduced in the inflationary potential, they become dominant over the non-perturbative contributions as soon as the inflaton is displaced from its minimum. Therefore, the inflationary potential takes a new inverse-power law behavior. We show that slow-roll inflation is not ultimately spoiled and focus on the post-inflationary history in different scenarios of microscopic Standard Model (SM) realization. Each of them gives precise predictions for the spectral index and the tensor-to-scalar ratio in agreement with CMB data \cite{Planck:2018vyg}. The model also predicts an amount of dark radiation potentially observable by next-generation CMB experiments.}

\FullConference{2nd Training School and General Meeting of the COST Action COSMIC WISPers  (CA21106) (COSMICWISPers2024)\\
 10-14 June 2024 and 3-6 September 2024\\
Ljubljana (Slovenia) and Istanbul (Turkey)\\}

\begin{document}

\maketitle

\section{Inflating with \Kahler moduli}
\begin{figure}[H]
    \centering
    \includegraphics[width=0.5\linewidth, height=0.27\linewidth]{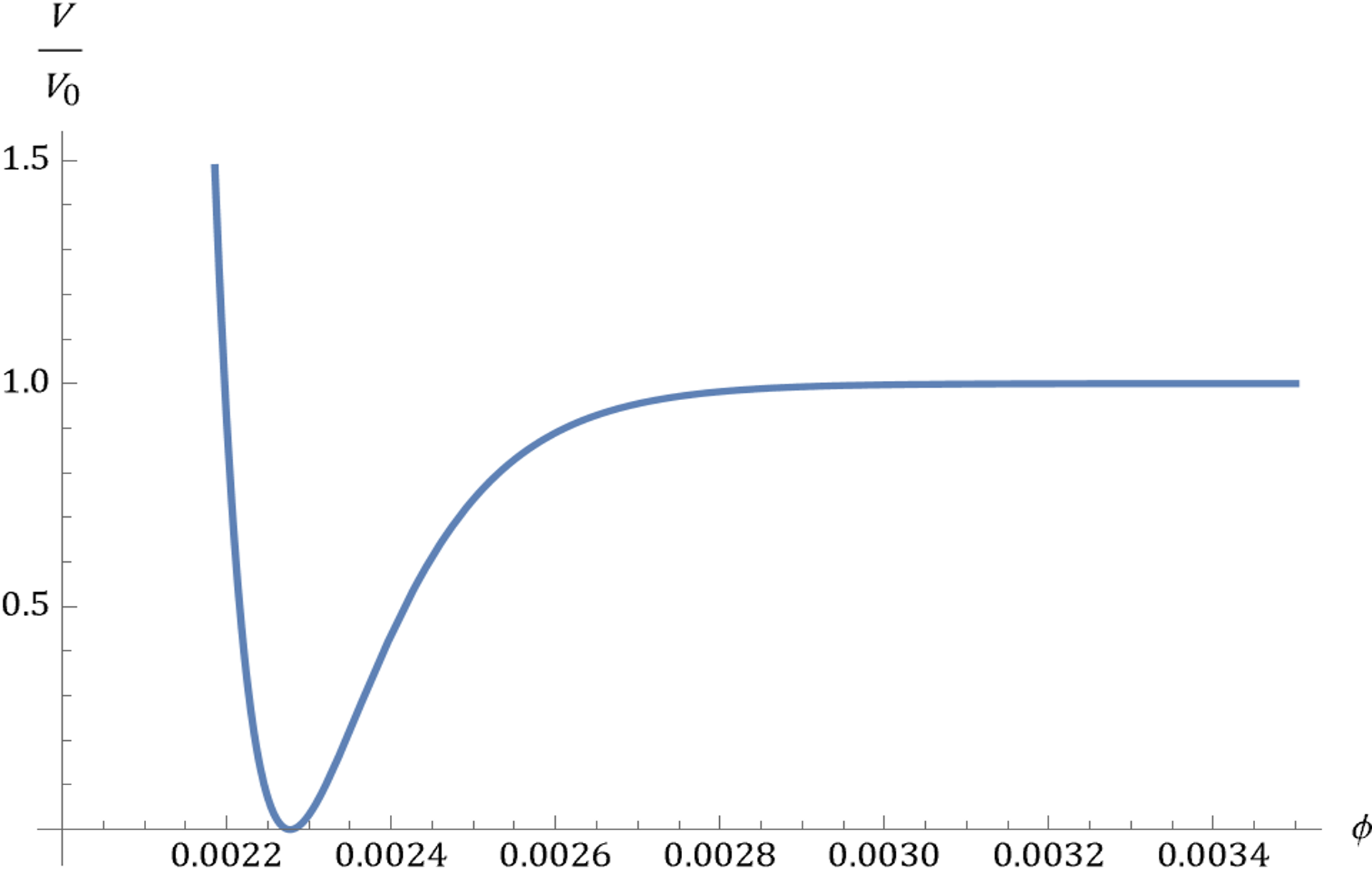}
    \caption{\footnotesize Potential of non-perturbative blow-up inflation with $\V = 10^6$, $a_\phi = 2 \pi$, $\tilde C_\phi = \left(\frac{3}{4}\right)^{2/3} 8 \pi$.}
    \label{fig:NP-Potential}
\end{figure}

Upon compactifying Type IIB string theory on a Calabi-Yau (CY) orientifold, one gets a 4d Effective Field Theory (EFT) featuring $\N = 1$ SUSY. As a side product, one gets a huge number of \textit{moduli}, 4d scalar fields with leading-order vanishing potential. In particular, the \textit{\Kahler moduli} sector of the 4d theory exhibits a classically flat potential, which is only lifted by quantum effects. This leading-order flatness makes a \Kahler modulus a great candidate for a slow-roll \textit{inflaton} field.
A natural example of this is the model we will call \textit{non-perturbative blow-up inflation} \cite{Conlon:2005jm}.\\ The Large Volume Scenario (LVS) balances $\O(\alpha^{'3})$ corrections to the \Kahler potential $K$ against non-perturbative corrections to the superpotential $W$ of the $\N=1$ EFT to stabilize the moduli. Let us consider a CY with a volume of the form:
\begin{equation}
    \V = \tau_b^{3/2}- \tau_s^{3/2}- \tau_\phi^{3/2} \quad \text{with } \tau_b^{3/2}\gg \tau_i^{3/2}\quad i = s, \phi
\end{equation}
where the $\tau$'s are moduli controlling the volume of 4-cycles called \textit{blow-up} modes, as they resolve localized singularities in the CY geometry. Trading $\tau_b$ for the overall volume, we get an F-term scalar potential of the form:
\begin{equation}\label{eq:LVS potential}
    V_{\rm LVS} = \tilde{V} \left[\sum_{i=s, \phi}\left(B_i \frac{\sqrt{\tau_i}\,e^{-2 a_i \tau_i}}{\V}- C_i \frac{ \tau_i \, e^{-a_i \tau_i}}{\V^2}\right) + \frac{3 \xi}{4 g_s ^{3/2} \V^3} + \delta V_{\rm up}\right]
\end{equation}
where $B_i, C_i, \xi $ are constants depending on the details of the compactification and $\delta V_{\rm up}$  uplifts the minimum to Minkowski. This ensures that the potential has a non-SUSY minimum with:
\begin{equation}
    \expval{a_i \tau_i} \sim \xi^{2/3}g_s^{-1}, \qquad \expval{\V} \sim e^{a_i \tau_i}. 
\end{equation}
Suppose that $\tau_\phi$ is displaced from its minimum towards larger field values while keeping the other moduli fixed. The resulting potential is shown in Fig \ref{fig:NP-Potential}. In the inflationary regime, $\tau_\phi \gg \expval{\tau_\phi}$, we can drop the doubly-exponentially suppressed term:
\begin{equation}
 V(\tau_\phi) \simeq V_0 \left[1- C_\phi \V \tau_\phi\, e^{-a_\phi \tau_\phi} \right]   
\end{equation}
and, upon canonical normalization $\tau_\phi = (3 \V/4)^{2/3} \phi^{4/3}$, we get:
\begin{equation}\label{eq:non-perturbative blow-up potential}
    V(\phi) \simeq V_0 \left[1- \tilde C_\phi \V^{5/3} \phi^{4/3} \exp[-\tilde a_\phi \V^{2/3} \phi^{4/3}]\right]
\end{equation}
where all the constants due to the canonical normalization have been reabsorbed in $\tilde C_\phi$ and $\tilde a_\phi$, and $V_0 \simeq \tilde V \V^{-3}$. 
In Fig \ref{fig:NP-Potential} one can see that this potential exhibits an exponentially flat plateau already for small displacements of $\phi$, making it a perfect slow-roll potential.

\section{String Loops and Loop Blow-Up Inflation}

\begin{figure}[H]
    \centering
    \includegraphics[width=0.5\linewidth,  
    height=0.25 \linewidth]{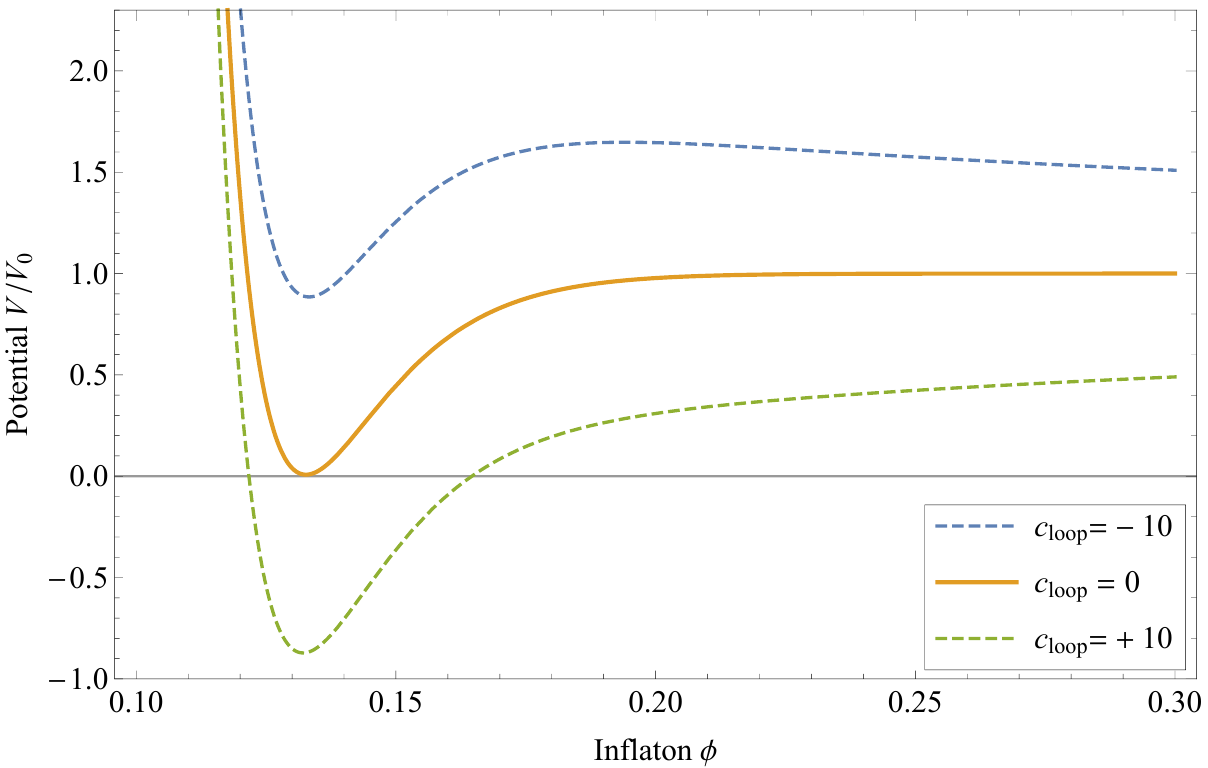}
    \caption{\footnotesize Loop Blow-Up potential for different values of $\cloop$, setting $\V \simeq 10^3$, $a_\phi=C_\phi =1$. The uplift here is tuned by hand to distinguish the three cases.}
    \label{fig:P-Potential}
\end{figure}

The extreme  non-perturbatuive flatness of this potential makes it very susceptible to radiative corrections in the form of string-loop corrections, as already noted in \cite{Conlon:2005jm}.
As of today, we still lack a precise calculation of string-loop corrections to the EFT on a CY background. However, a computation of 1-open-string loop corrections has been performed in a toroidal orientifold background in \cite{Berg:2005ja}, and later generalized to a conjecture for CY orientifolds in \cite{Berg:2007wt}. Moreover, loop corrections have a precise understanding from the EFT point of view as contributions both to the 2-point functions and to the Coleman-Weinberg vacuum potential \cite{Cicoli:2007xp,  Gao:2022uop}.
For a blow-up mode $\tau$ the leading corrections to the scalar potential go as:
\begin{equation}\label{eq:loop correction}
    \delta V (\tau) \simeq \frac{\cloop}{\V^3 \sqrt{\tau}}
\end{equation}
where $\cloop$ is a loop factor which in principle can depend on the complex structure moduli; at this stage, however, we take it as a constant. 
Inserting this correction in the potential \eqref{eq:LVS potential} yields a potential as shown in Fig. \ref{fig:P-Potential}. 

Clearly, $\cloop<0$ leads to a runaway, so we stick to $\cloop >0$. For $\cloop \gtrsim 10^{-6}$, the loop contribution dominates as soon as the inflaton is displaced from its minimum. Then the inflationary potential in terms of $\phi$ becomes:
\begin{equation}\label{eq:loop blow-up potential}
    V(\phi) \simeq V_0\left[1- \frac{b\, \cloop}{\V^{1/3} \phi^{2/3}}\right].
\end{equation}
where the constant $b$ contains the canonical normalization terms. This is a completely different inflationary potential from \eqref{eq:non-perturbative blow-up potential} and thus has different inflationary dynamics.

\section{Inflationary and Post-Inflationary Dynamics}

Using \eqref{eq:loop blow-up potential}, we can retrieve the number of e-folds $N_e$, the spectral index $n_s$, the tensor-to-scalar ratio $r$ and the amplitude of the scalar perturbations during inflation $A_s$:

\begin{align}
     N_e & \simeq \frac{9}{16} \frac{\V^{1/3} \phi_*^{8/3}}{b \cloop}\\
     n_s & \simeq 1- \frac{20}{9} \frac{b \cloop}{\V^{1/3} \phi_*^{8/3}}
\end{align}
\begin{align}
    r &  \simeq \frac{32}{9} \frac{(b \cloop)^2}{\V^{2/3} \phi_*^{10/3}}\\
    A_s & \simeq \frac{3 V_0}{16 \pi^2} \frac{\V^{2/3} \phi_*^{10/3}}{(b \cloop)^2}
    \end{align}
where $\phi_e$ and $\phi_*$  are the values of the inflaton field at the end of inflation and at horizon exit respectively. Setting $\cloop = 1/(16 \pi^2)$ and imposing  $A_s = 2 \times 10^{-9}$ \cite{Planck:2018vyg} we can get a relation between $\phi_*$, $\V$ and the number of e-folds:
\begin{equation}
    \phi_* = 0.06 \, N_e^{7/22}, \qquad \V = 1743 \, N_e^{5/11}
\end{equation}
Therefore, to get a precise prediction from the model we need to determine the number of e-folds from post-inflationary dynamics. To do so, we adapt to our specific case a relation from modular cosmology \cite{Dutta:2014tya}:
\begin{equation}\label{eq:Ne moduli}
    N_e \simeq 57 + \frac{1}{4} \ln r - \frac{1}{4} N_\phi -\frac{1}{4}N_\chi
\end{equation}
where $N_\phi$ and $N_\chi$ are the number of e-folds of inflaton and volume domination after the end of inflation. These in turn depend on the microscopic realization of the Standard Model (SM) in the compact dimensions, i.e.the brane set-up. Moreover, the last dominating modulus to decay sources an amount of Dark Radiation (DR) depending on its branching ratio into hidden sector light modes (e.g. volume mode axions). This is quantified in terms of the effective number of extra relativistic species $\Delta N_{\rm eff}$, bounded by \cite{Planck:2018vyg} to be $\Delta N_{\rm eff} \lesssim 0.5$ at $95 \%$ CL. 

We distinguish 3 scenarios of post-infaltionary evolution based on the brane setup determining the decay rate of the moduli:
\begin{enumerate}
    \item \textit{Scenario I}: SM is realized on D7-branes and the inflaton 4-cycle is wrapped by a D7-brane stack supporting a hidden sector
    \item \textit{Scenario II}: SM is realized on D7-branes and the inflaton 4-cycle is \textit{not} wrapped by  D7-branes
    \item \textit{Scenario III}: The SM is realized on D3-branes at singularity. 
\end{enumerate}
In each of these scenarios the post-inflationary history of the Universe has a slightly different play-out, but there are many common features. At the end of inflation, the inflaton field starts oscillating around its minimum, behaving as a cold matter condensate. The duration of this period depends on the Scenario, and lasts until the inflaton decays, $1 \lesssim N_\phi \lesssim 11$.
The decay products of $\phi$ dominate the energy density, redshifting as radiation. Meanwhile, the volume mode, whose canonically normalized counterpart we call $\chi$, oscillates around its minimum due to a slight misalignment during inflation \cite{Cicoli:2016olq}, and redshifts as matter. Therefore, at some point it will become dominant for $0 \leq N_\chi \lesssim 10.5$, until its decay. $N_\chi = 0$ corresponds to Scenario II, where $\chi$ decays \textit{before} matter-radiation equality. Using \eqref{eq:Ne moduli} we get the number of efolds of inflation:
\begin{equation}
    51.5 \lesssim N_e \lesssim 53 \implies \V \simeq \O(10^4) \quad \text{and}\quad  \phi_* \simeq 0.2\, .
\end{equation}
Thus we can have a precise prediction for the spectral index and the tensor-to-scalar ratio:
\begin{equation}
    n_s \simeq 0.976 \quad \text{and} \quad r \simeq 2 \times 10^{-5}. 
\end{equation}
which are within observational bounds \cite{Planck:2018vyg}. One of the most intriguing predictions of this model is the abundance of DR $\Delta N_{\rm eff}$. Although in Scenario I this ends up being negligibly small, in Scenario III it may get to $\Delta N_{\rm eff} \simeq 0.36$, which is well within reach of next-generation searches and experiments. 

\section{Conclusions and Outlook}

In this talk, we have presented a new inflationary model which solves the $\eta$-problem of \textit{non-perturbative blow-up inflation}. The inflationary potential is given by perturbative corrections to the potential of a blow-up modulus and inflation is achieved in a region of field-space with full control over the EFT.
Moreover, the values for $n_s$ and $r$ for this model are within current experimental bounds, and, most importantly, the prediction for $\Delta N_{\rm eff}$ is within reach of next-generation experiments.

Since this model combines perturbative and non-perturbative corrections in the same potential, much research still needs to be done.
First, the role of the axions in the phenomenology of inflation must be established. Moreover, the robustness of the prediction of this model under higher loop corrections and higher derivative corrections still needs to be assessed. Finally, realizing this model with a fully perturbative stabilization of the inflaton 4-cycle would be interesting, as it could radically modify the post-inflationary dynamics and the phenomenological predictions. 

\section*{Acknowledgements}
I thank S. Bansal, M. Cicoli, A. Hebecker and R. Kuespert for their collaboration in this project. I would also like to thank M. Cicoli for enlightening discussions about the present work. 
This article is based upon work from COST Action COSMIC WISPers CA21106,
supported by COST (European Cooperation in Science and Technology).

\end{document}